\documentclass[12pt, preprint]{aastex}

\newcommand{\spi}{{\em Spitzer}}
\newcommand{\msx}{{\em MSX}}
\newcommand{\mum}{{\micron}}
\newcommand{\rstar}{\ifmmode{\rm R_*}\else{R$_*$}\fi}
\newcommand{\msun}{\ifmmode{\rm M_{\odot}}\else{M$_{\odot}$}\fi}
\newcommand{\lsun}{\ifmmode{\rm L_{\odot}}\else{L$_{\odot}$}\fi}

\begin{document}

\title{Circumstellar Structure around Evolved Stars in the Cygnus-X Star 
Formation Region}

\author{Kathleen E. Kraemer\altaffilmark{1}, Joseph L. Hora\altaffilmark{2}, 
Michael P. Egan\altaffilmark{3}, Joseph 
Adams\altaffilmark{4}, Lori E. Allen\altaffilmark{5}, Sylvain 
Bontemps\altaffilmark{6}, Sean J. Carey\altaffilmark{7}, Giovanni G.
Fazio\altaffilmark{2}, Robert Gutermuth\altaffilmark{8,9}, Eric
Keto\altaffilmark{2}, Xavier P. Koenig\altaffilmark{2}, S. Thomas Megeath\altaffilmark{10}, Donald R. 
Mizuno\altaffilmark{11}, Frederique Motte\altaffilmark{12}, Stephan D. 
Price\altaffilmark{1}, Nicola
Schneider\altaffilmark{12}, Robert Simon\altaffilmark{13}, \and
Howard Smith\altaffilmark{2}}

\begin{abstract}

We present observations of newly discovered 24 \mum\ circumstellar structures 
detected with the Multiband Imaging Photometer for {\em Spitzer} (MIPS) around 
three 
evolved stars in the Cygnus-X star forming region. One of the objects, 
BD+43 3710, has a bipolar nebula, possibly due to an outflow or a torus of 
material. A second, HBHA 4202-22, a Wolf-Rayet candidate, shows a circular 
shell of 24 \mum\ emission suggestive of either a limb-brightened shell or 
disk seen face-on. No diffuse emission was detected around either of these 
two objects in the {\em Spitzer} 3.6-8 \mum\ Infrared Array Camera 
(IRAC) bands. 
The third object is the luminous blue variable candidate G79.29+0.46. We 
resolved the 
previously known inner ring in all four IRAC bands. The 24 \mum\ emission 
from the inner ring extends $\sim1\farcm2$ beyond the shorter wavelength 
emission, well beyond what can be attributed to the difference in resolutions
between MIPS and IRAC. Additionally, we have discovered an outer ring of 
24 \mum\
emission, possibly due to an earlier episode of mass loss. For the two shell
stars, we present the results of radiative transfer models, constraining the 
stellar and dust shell parameters. The shells are composed of amorphous carbon
grains, plus polycyclic aromatic hydrocarbons in the case of G79.29+0.46. 
Both G79.29+0.46 and HBHA 4202-22 lie behind the main Cygnus-X cloud. Although 
G79.29+0.46 may simply be on the far side of the cloud, HBHA 4202-22 is 
unrelated to the Cygnus-X star formation region.
\end{abstract}

\altaffiltext{1}{Air Force Research Laboratory, Space Vehicles Directorate, 
29 Randolph Road, Hanscom AFB, MA 01731; afrl.rvb.pa@hanscom.af.mil} 
\altaffiltext{2}{Harvard-Smithsonian Center for Astrophysics, 60 Garden Street,
 MS 65, Cambridge, MA 02138; jhora@cfa.harvard.edu, gfazio@cfa.harvard.edu, 
eketo@cfa.harvard.edu, xkoenig@cfa.harvard.edu, hsmith@cfa.harvard.edu}
\altaffiltext{3}{National Geo-Spatial Intelligence Agency, DN-11, 12310 
Sunrise Valley Dr., Reston, VA 20191; michael.p.egan@nga.mil}
\altaffiltext{4}{Cornell University; jdadams@astro.cornell.edu}
\altaffiltext{5}{National Optical Astronomy Observatory; lallen@noao.edu}
\altaffiltext{6}{Observatoire de Bordeaux; Sylvain.Bontemps@obs.u-bordeaux1.fr}
\altaffiltext{7}{\spi\ Science Center, Caltech, MS 220-6, Pasadena, CA 91125; 
carey@ipac.caltech.edu}
\altaffiltext{8}{Five College Astronomy Department, Smith College, Northampton, MA 01063; rgutermu@smith.edu}
\altaffiltext{9}{Department of Astronomy, University of Massachusetts, Amherst, MA 01003}
\altaffiltext{10}{Department of Physics and Astronomy, University of Toledo, 
Toledo, OH 43606; meageath@astro1.panet.utoledo.edu}
\altaffiltext{11}{Institute for Scientific Research, Boston College, Chestnut 
Hill, MA 02467-3800; afrl.rvb.pa@hanscom.af.mil}
\altaffiltext{12}{Commissariat a l'energie atomique, Saclay; 
motte@discovery.saclay.cea.fr, nschneid@cea.fr}
\altaffiltext{13}{Universitat zu Koln; simonr@ph1.uni-koeln.de}

\section{Introduction \label{sec.intro}}

Cygnus-X is arguably the biggest and brightest massive 
star formation region within 2 kpc of the Sun (d$\sim$1.7 kpc, Schneider et al.
2006, 2007). It contains several hundred
HII regions, numerous OB associations with over a thousand OB stars, and 
thousands of low mass stars. We have performed an unbiased survey of 
$\sim$24 sq degrees
in Cygnus-X with the Infrared Array Camera (IRAC, 
Fazio et al. 2004) and the Multiband Imaging Photometer for \spi\ (MIPS, 
Rieke et al. 2004) on the {\em Spitzer Space Telescope} (Werner et al. 2004).
The primary objectives of this \spi\ Legacy project were to 
study the evolution of high mass 
protostars, clustering of both high and low mass stars in the complex, as 
well as several other goals described by Hora et al. (2008b). Follow-up 
spectroscopy with \spi's Infrared Spectrograph (IRS; Houck et al. 2004) was 
obtained for several dozen promising sources in the MIPS and IRAC survey 
regions.

In addition to the numerous young stars that are the primary focus of the 
Legacy 
project, there are dozens of evolved objects in the survey region, from 
Miras and Cepheids to Wolf-Rayet and carbon stars to planetary nebulae and
supernova remnants. After the initial discovery
of a bipolar nebula at 24 \mum\ around a carbon star during the data processing
quality control effort, a systematic search for additional circumstellar 
nebulae was made. SIMBAD was used to compile positional lists of evolved 
objects that lie
within the bounds of the 24 \mum\ observations. The 24 \mum\ images were then
inspected for the presence of point sources and associated nebulosity, if any.
More than 80\% (146) of the 191 known evolved objects are detected 
at 24 \mum\ as point sources (Table 1). These types of objects can be bright
in the infrared for a variety of reasons such as emission from dust or 
forbidden lines. This, of course, is an incomplete
survey of the evolved stars as it does not include the faint population
of asymptotic giant branch (AGB) stars detected with \spi\ (also known as
contaminants in the young stellar object studies). Be that as it may, 
only three evolved stars in the Cygnus-X region were distinguished from the
vast majority by exhibiting extended 24 \mum\ emission. The emission
around two of these stars, which are listed in Table 2, was first reported
by Kraemer et al. (2009). Here, we discuss the morphological characteristics
of the circumstellar structures as well as the spectral properties of
the dust and the central exciting objects.

\section{Observations and Data Reduction\label{sec.obs}}

The Cygnus-X Legacy Survey (Hora et al. 2008b) observed a 
$\sim7\arcdeg\times7\arcdeg$ region centered
on $(l, b) \sim (80\arcdeg, 0\fdg3)$ with \spi's IRAC (all 4 bands) and 
MIPS (24 and 70 \mum). The MIPS observations followed the MIPSGAL survey
strategy (Carey et al. 2009), using the fast-scanning 
mode with 3 sec integrations and 148\arcsec\ between legs. This results in 
$\sim10$ times redundancy per pixel or $\sim30$ sec total integration time.
We use the 24 \mum\ data processing pipeline optimized for MIPSGAL, which 
is based on the \spi\ Science Center (SSC) pipeline ($\sim$S16.1). It includes 
the standard ``read-2'', row-droop, and droop corrections, dark subtraction, 
flat fielding, and a modified linearization. The optimization involves better 
artifact mitigation such as overlap, jailbar, and latent corrections; 
additional details are described by Mizuno et al. (2008) and Carey et al. 
(2009). The 70 \mum\ data are not considered here and are thus not described. 
For completeness, we note that BD+43 3710 was outside the 70 \mum\ coverage, 
HBHA 4202-22 is not detected in the post-bcd images, and the inner ring of 
G79.29+0.46 is apparent in the post-bcd images but the new outer ring is not
(see Section 3.3). 

The Cygnus-X survey measurements with IRAC consisted of 
$\sim1\fdg1\times1\fdg1$ tiles 
taken in high dynamic range mode with 3$\times$12 sec integrations. The 
IRAC data were processed using the SSC pipeline S16.1 products with
 artifact mitigation performed for jailbar, pulldown, muxbleed, and
banding effects from bright sources.  Mosaics were created with the WCSmosaic
package which performs additional corrections detailed by Gutermuth et al. 
(2008). Photometry was extracted using the ``PhotVis'' package (Gutermuth et 
al. 
2008) for the IRAC data and the MIPSGAL point source pipeline (Shenoy et al., 
in 
preparation) for the 24 \mum\ data.

BD+43 3710, discovered as being extended in the MIPS data taken in 2007 
December, was outside the IRAC coverage and so was observed in a 
separately designed Astronomical Observation Request (AOR). The same data 
reduction process was applied as 
with the primary IRAC observations.  
HBHA 4202-22 was at the edge of the IRAC coverage region and was only 
observed with the 3.6 and 5.8 \mum\ arrays, as the
circumstellar ring was not discovered in time to adjust the AORs or add an 
additional field as was done for BD+43 3710.

We also observed BD+43 3710 with the IRS, 
targeting the central star and one position in each lobe of the 24 \mum\ 
bipolar 
nebula. The observations used all four low resolution modes and the data 
(Pipeline S18.0.2) were 
reduced using the \spi\
IRS Custom Extraction (SPICE) software, version 2.1.2, provided by the SSC. 
For the central 
star observations,
 we subtracted the alternate order for the Short-Low (SL) data and the 
alternate 
nod for the Long-Low (LL) data prior to performing the point source extraction. 
After the spectra
were extracted with SPICE, they were trimmed to get rid of bad data at the 
extremes of the wavelength ranges and averaged together. For the positions
in the lobes, we used the full slit extraction, which includes corrections to 
the calibration due to the aperture size and extended nature of the target. 
There were no compact sources apparent in either the SL or LL slit for the 
lobe observations; no 
background subtraction was performed.
The luminous blue variable (LBV) candidate G79.29+0.46 was observed
with IRS in high resolution mode by another group and results are presented by
Umana et al. (in preparation).

Photometric measurements from this work and from the literature are summarized
in Table 3 for each of the three stars. BVR data are from the Naval Observatory
Merged Astrometric Dataset (NOMAD; Zacharias et al. 2004); JHK data are from
the Two Micron All-Sky Survey (Cutri et al. 2003); mid-infrared data are from our
\spi\ measurements. The uncertainties for the \spi\ data are 
the extraction-based uncertainties combined with the absolute 
uncertainties for IRAC (2\%, Hora et al. 2008a) and MIPS (4\%, Engelbracht et
al. 2007). Those for 2MASS are from Cutri et al. (2003; see also Skrutskie et al. 2006); the BVR uncertainties
are estimated at 4\% as they are not listed in NOMAD.

\section{Results  and Discussion\label{sec.results}}

\subsection{BD+43 3710}

\begin{figure}
\plotone{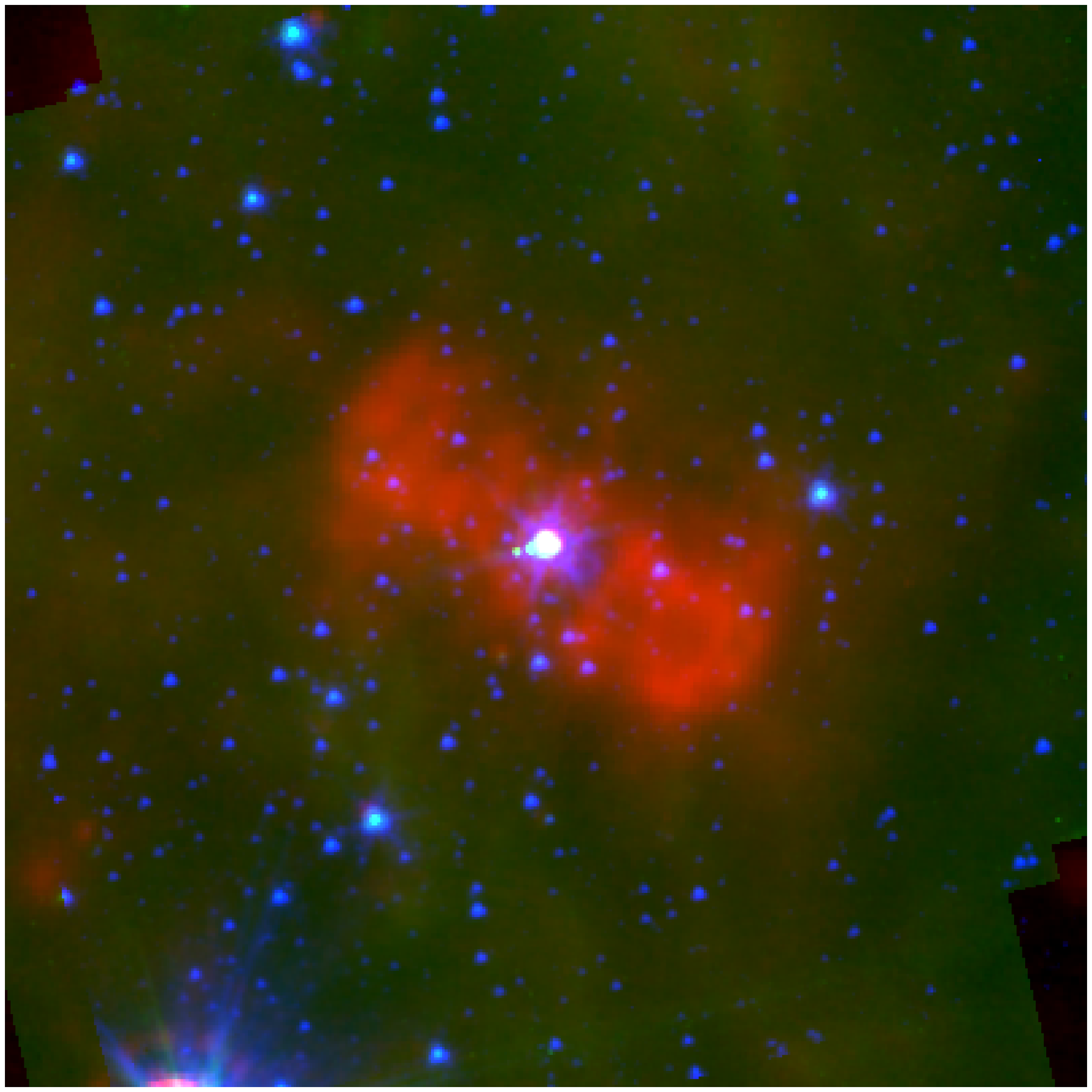}
\caption{Three color image of BD+43 3710. Red: 24 \mum; green: 8 \mum; blue:
3.6 \mum. The image is $5\farcm78\times5\farcm78$, or $\sim2.9\times2.9$ pc,
 assuming a distance of 1.7 kpc.}
\label{fig.bd_3c}
\end{figure}

\begin{figure}
\plotone{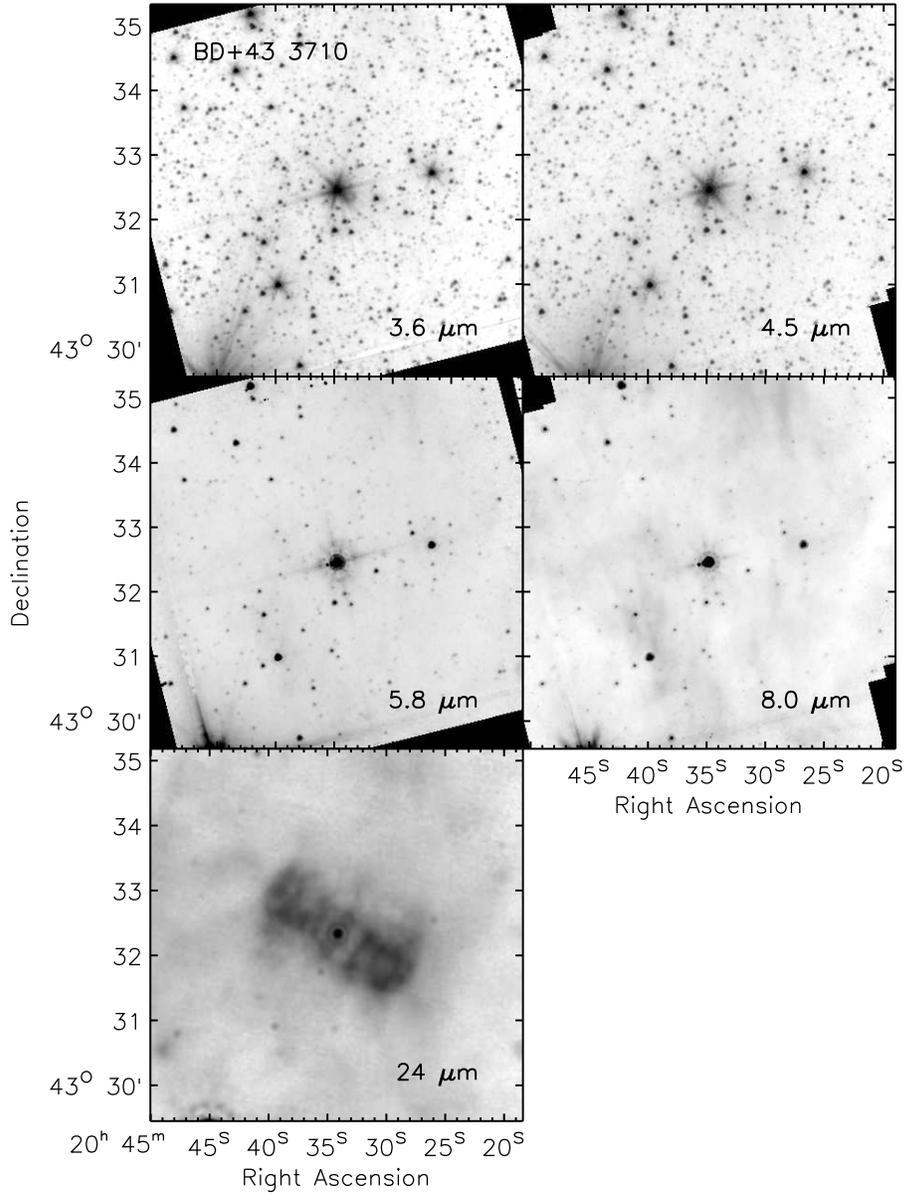}
\caption{Individual images of BD+43 3710 in the IRAC bands (top two rows) and
in MIPS 24 \mum\ (bottom image). Gray scales are logarithmic.} 
\label{fig.bd_5}
\end{figure}

BD+43 3710, located in the extreme northeast corner of the Cygnus-X survey 
region, is a little known carbon star candidate classified as spectral
type R: by Nassau \& Blanco (1954). It shows a remarkable bipolar structure at 
24 \micron\ that is completely absent at the shorter wavelengths (it is also 
outside the 70 \micron\ MIPS coverage). Figure \ref{fig.bd_3c} shows a 
three-color image of BD+43 3710 combining the 3.6 \mum, 8 
\mum, and 24 \mum\ data while Figure \ref{fig.bd_5} shows the individual 
images from all
four IRAC bands plus the MIPS 24 \mum\ image.  The 24 \mum\ emission is
roughly  $2\farcm7\times0\farcm9$ or $\sim1.34\times0.45$ pc, assuming the
Cygnus-X distance of 1.7 kpc. 

\begin{figure}
\epsscale{0.5}
\plotone{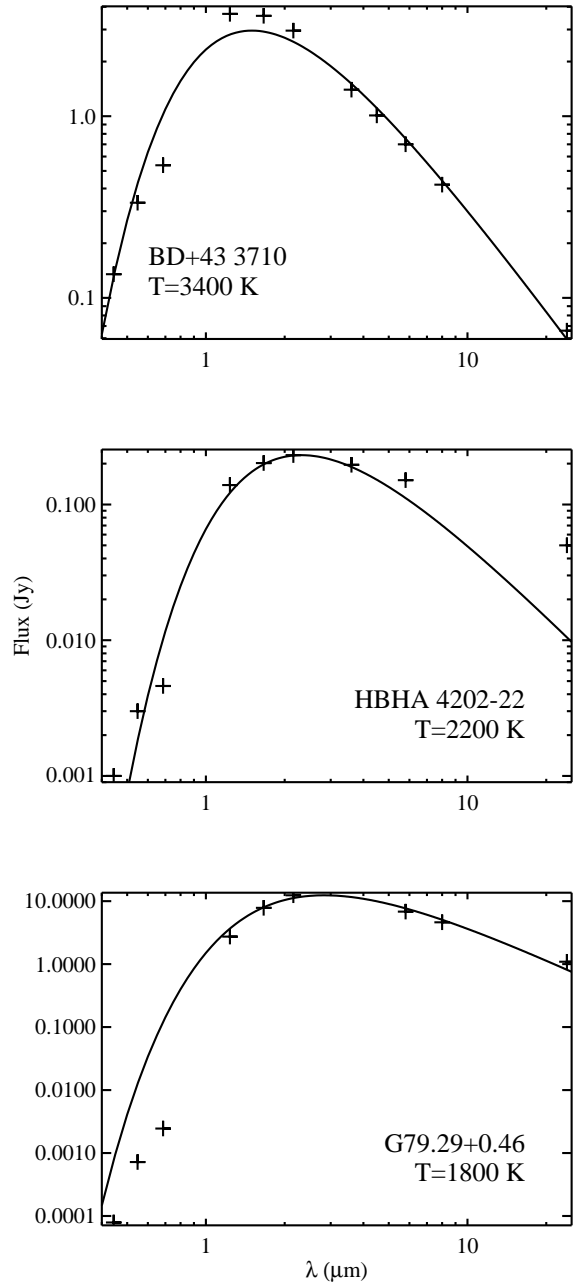}
\caption{SEDs for (top) BD+43 3710, (middle) HBHA 4202-22, and (bottom) 
G79.29+0.46. Temperatures for the best fit (minimal $\chi^2$) graybody curve
(smooth curves) are given in the lower left of each panel. Pluses correspond 
to the photometry from Table 3. Uncertainties are smaller than the symbol size.}
\epsscale{1.0}
\label{fig.sed}
\end{figure}

The central star is in the Bonner Durchmusterung (Argelander 1903), Tycho 
(H\o g et al. 2000), 2MASS (Cutri et al. 2003), and 
{\em Midcourse Space Experiment} (\msx; Egan et al. 2003) point source 
catalogs. However, no references discussing its physical properties could be 
found in the literature. The visual magnitude did change from 
a reported 9.2 mag in the Bonner Durchmusterung (1903) to 10.1 mag  in the 
early 1990s as measured by Tycho. Other than indicating variability in this 
object, not an unusual feature in a  carbon star, such a sparse sampling of 
the  temporal baseline (albeit long) does not allow us to say anything 
meaningful about the pulsation properties of the star. Figure \ref{fig.sed}
 shows
the spectral energy distribution for BD+43 3710, as well as the other two 
sources. The smooth curves are Planck curve fits to the available photometry.
A single temperature graybody does not well represent the emission
from any of these sources, although the BVR data are probably suppressed by an
unknown amount of extinction.

\begin{figure}
\plottwo{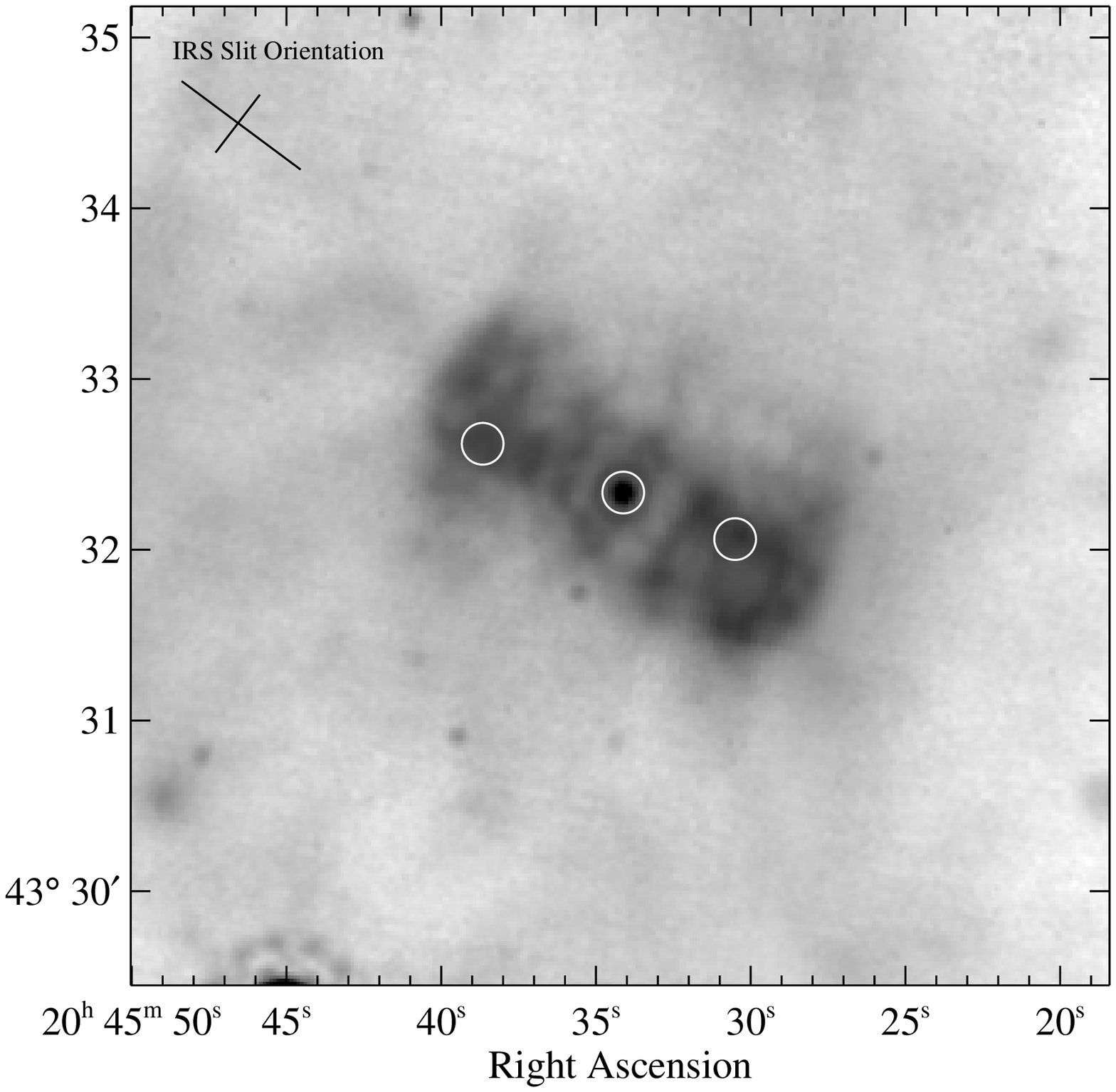}{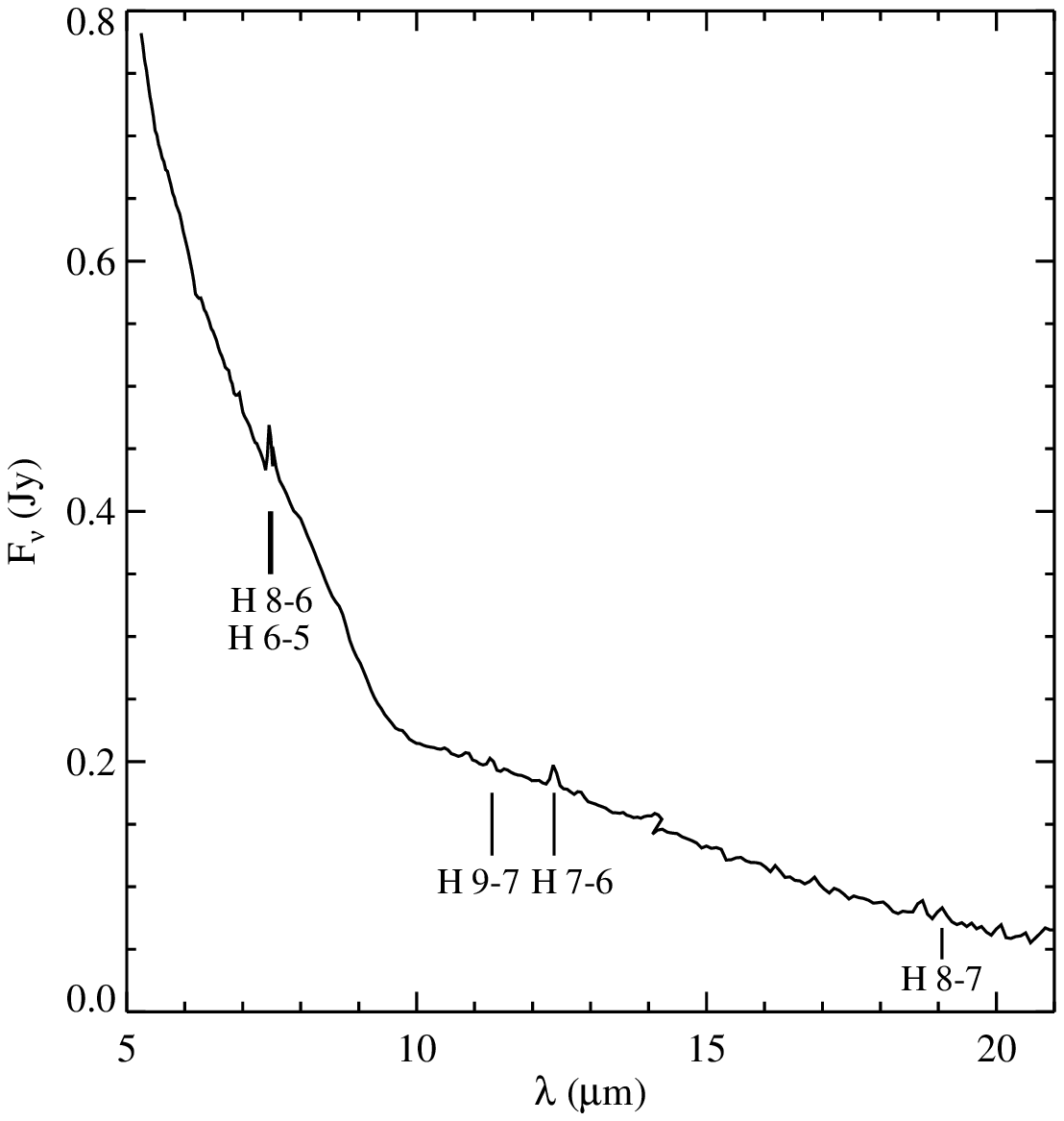}
\caption{(left) Positions (circles) observed with IRS in BD+43 3710, overlaid 
on the 24 \micron\ data (grayscale). The orientation of the SL and LL slits is
shown in the upper left corner. 
(right) IRS 5.25--21 \micron\ spectrum of BD+43 3710. A handful of hydrogen
recombination lines are detected in emission and are labeled on the plot. The
feature slightly shortward of H 8-7 is only present in one nod and so is 
probably spurious.
}
\label{fig.irs_star}
\end{figure}

As mentioned above, we obtained low-resolution IRS data on BD+43 3710 and 
one position in each lobe. Figure \ref{fig.irs_star} (left) shows the
locations where the spectra were taken. The right side shows the resulting 
spectrum for the central star. The spectrum is largely featureless. In 
comparison to the naked stars in the ISOSWS atlas of Sloan et al. (2003), 
it is most similar to HD 19557, a carbon star of type R noted by Goebel et al.
(1983) as 
having particularly weak carbon features in visible and near-IR bands. 
In particular, the inflection in the IRS spectrum at 9 \micron\ is well-matched
in the SWS spectrum of HD 19557, although it is not apparent in most of the 
naked carbon stars in the SWS atlas (class 1.NC of Kraemer et al. 2002) or in
the other dust-free SWS spectral classes (1.N, 1.NO, 1.NE, and 1.NM). The IRS 
spectrum,
however, does not show the turnover at $\sim$5.5 \micron\ expected from 
C$_3$, CO, and CN absorption features (Goebel et al. 1978) typically observed 
in carbon stars (e.g. Aoki et al. 1998, Zijlstra et al. 2006). 
Several hydrogen recombination lines are detected in emission, as annotated 
in the figure. The strengths of the Pf $\alpha$, Hu $\alpha$, and HI 8-7
do not change between the raw data and the background-subtracted data, unlike
the typical nebular lines such as the [Ne II] feature at 12.8 \micron, which 
nearly vanishes after background subtraction. Thus, the hydrogen recombination 
lines are likely associated with BD+43 3710 and its nebula, and are not simply 
in the foreground or background cloud.  Although not typical 
for carbon stars, Balmer lines have been detected in emission in the carbon 
star UV Aur A (Herbig 2009). Thus, we find that
the carbon star spectral classification for BD+43 3710 is supported by the IRS 
spectrum but not definitively confirmed.

\begin{figure}
\plottwo{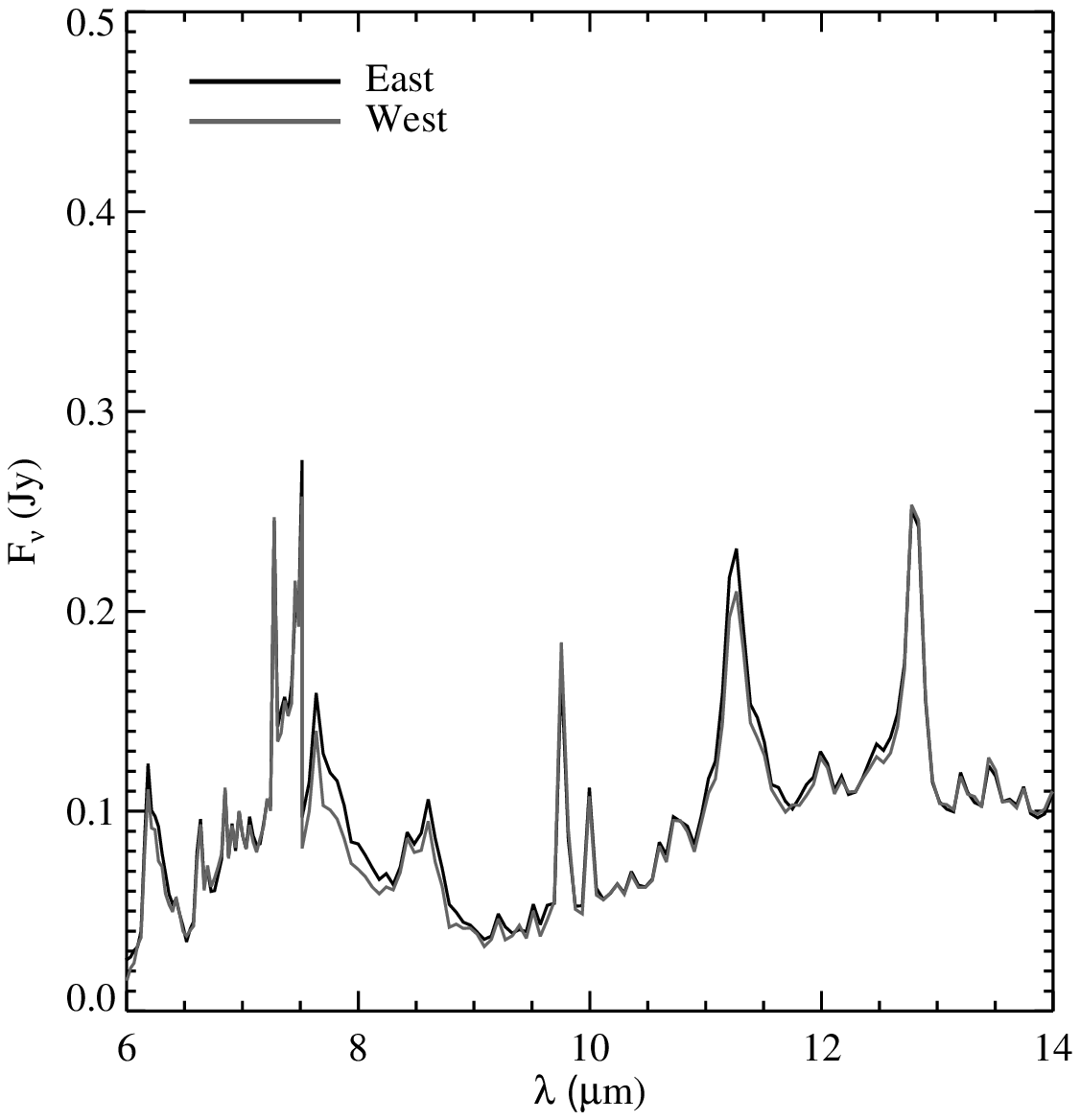}{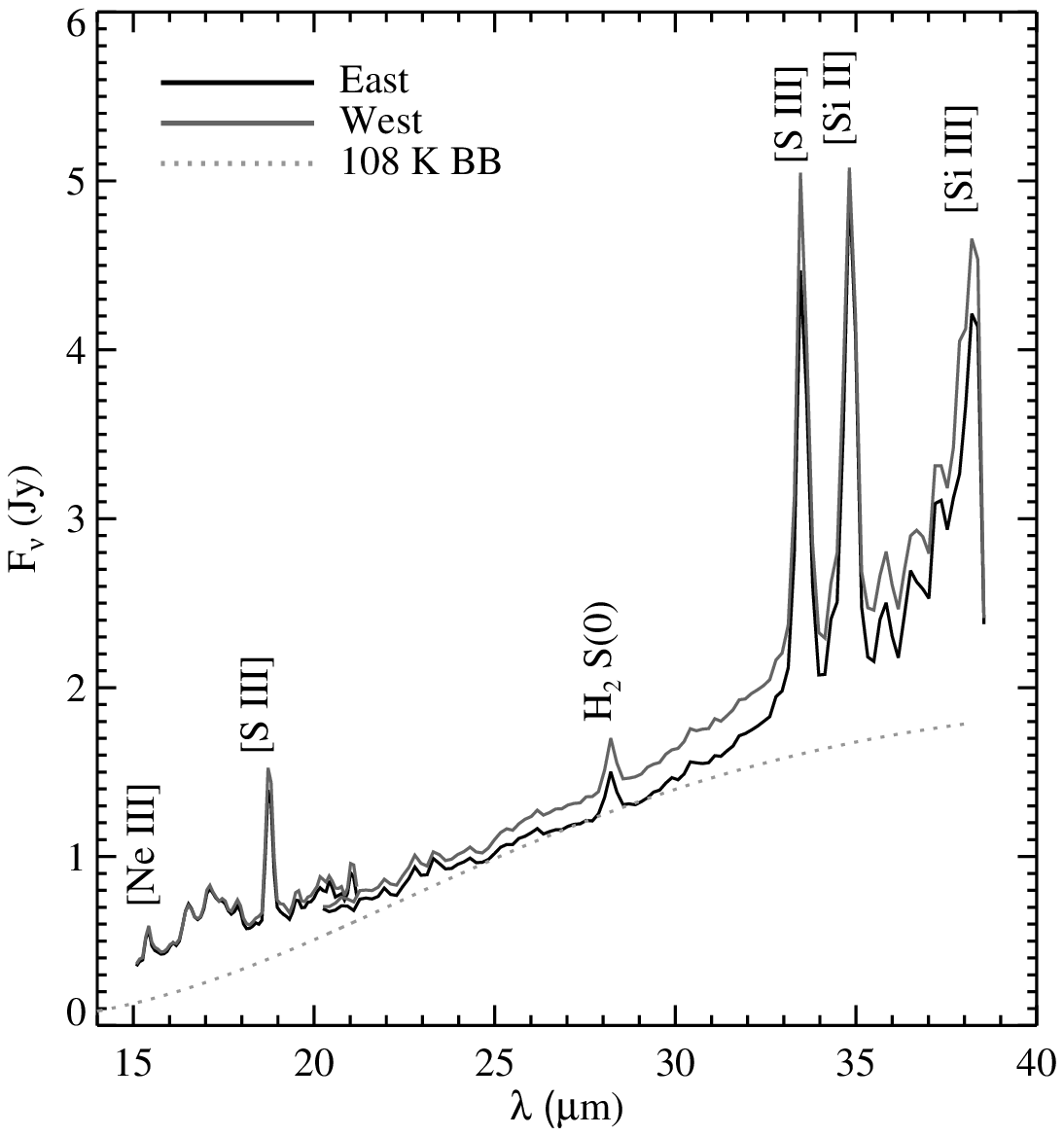}
\caption{(left) Short-low IRS spectra from the two lobes, east in black and
west in gray.  (right) Long-Low IRS spectra from the two lobes, east 
in black and west in gray. The labeled emission features likely belong to the
fore/background Cygnus-X cloud, not BD+43 3710, as discussed in the text. A 
108 K graybody is the ``best'' fit to the LL data for both lobes and is shown 
as the gray dotted line.}
\label{fig.irs_ew}
\end{figure}

The spectra from the two lobes, shown in Figure \ref{fig.irs_ew}, are quite 
similar to each other, with a cool dust spectrum that
rises with wavelength past the end of the IRS band and a few broad features 
and low-excitation fine structure lines. The SED of the western lobe rises a 
bit more steeply with wavelength than that of the eastern lobe, possibly
indicative of slightly warmer dust. Neither spectrum is consistent with a 
single temperature graybody, but the majority of the dust must be cooler than
$\sim$100 K since the SED peaks beyond 35 \mum. A 108 K graybody is the 
``best'' fit to the LL data for both lobes and is shown in the figure 
for comparison. As can be seen from the spectra, unlike in a small number of 
sources where the circumstellar emission in the MIPS 24 \mum\ band has been 
found to arise from a high excitation [O IV] line (e.g. Morris et al. 2004 
(a WN star), Morris et al. 2006 (a SNR candidate), Billot et al. 2009 (sources 
of unknown nature, possibly SNR or PNe candidates), Flagey et 
al. in preparation), or the $\sim26-30$ \mum\ MgS feature seen in some 
post-asymptotic giant branch stars and planetary nebulae (e.g. Forrest et al.
1981; Goebel \& Moseley 1985; Hony et al. 2002; Bernard-Salas et al. 2009), 
the nebular emission around BD+43 3710 arises from small dust grains. 
They are 
probably carbon-rich grains, dominated by amorphous carbon, given the 
lack of either a silicate emission feature around 9.7 \mum\ or silicon carbide
around 11 \mum. The broad features present in the spectra at 6.3, 7-8, 11.3, 
and 16-18 \mum,
typically attributed to polycyclic aromatic hydrocarbons (PAHs), are not 
likely to be from the BD+43 3710 nebula since there is no corresponding 
nebulosity in the IRAC 5.8 or 8.0 \mum\ image that can be readily associated
with the 24 \mum\ lobes (Fig. \ref{fig.bd_5}). The PAHs and low 
energy fine structure lines seen in Figure \ref{fig.irs_ew} are extremely 
common
in active star forming regions such as Cygnus-X, so those features likely 
arise from the fore/background Cygnus-X cloud.

\subsection{HBHA 4202-22}

\begin{figure}
\plotone{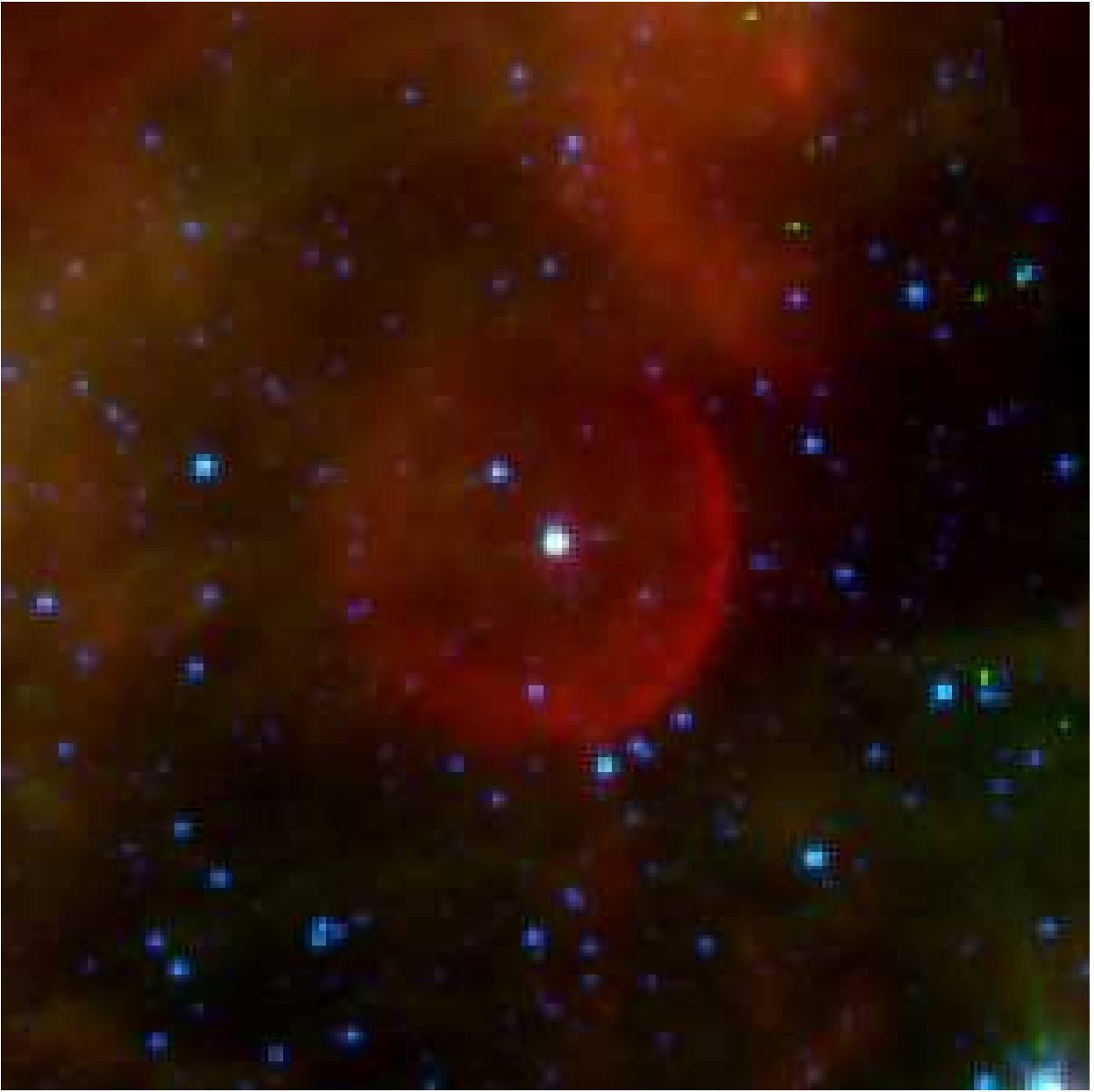}
\caption{Three color image of HBHA 4202-22. Red: 24 \mum; green: 5.8 \mum; 
blue: 3.6 \mum.  The image is $6\farcm02\times6\farcm02$ or $\sim$11.4 pc at 
a distance of 6.5 kpc (see text).
}\label{fig.hb_3c}
\end{figure}

\begin{figure}
\plotone{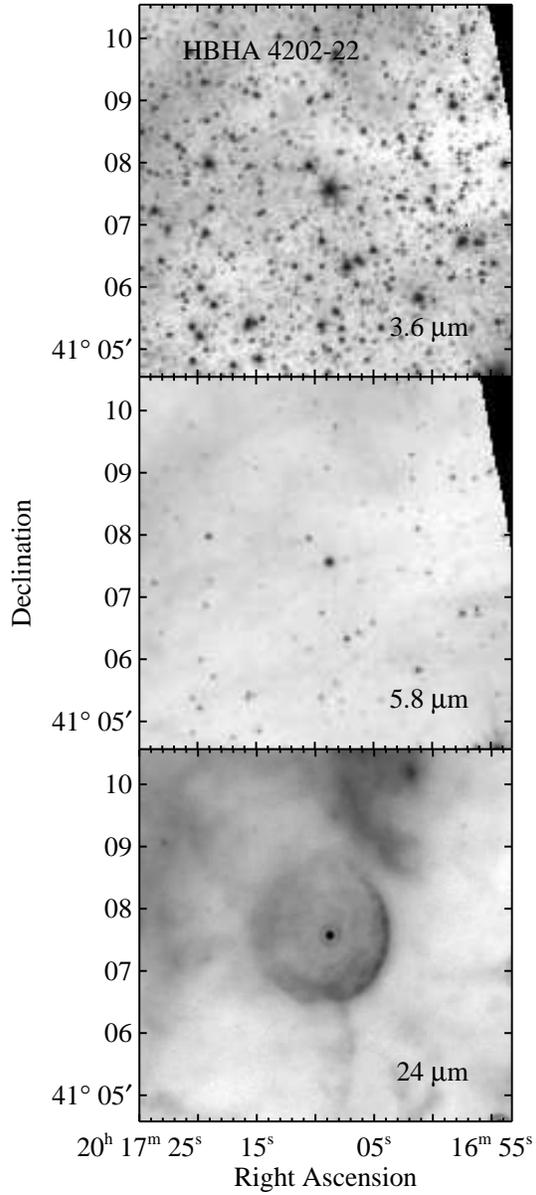}
\caption{Individual images for HBHA 4202-22. As with BD+43 3710, there is
no indication of extended emission associated with the star in the IRAC
bands (top and middle panels), in contrast to the circumstellar shell 
detected at 24 \micron\ (bottom).
}
\label{fig.hb_all}
\end{figure}

HBHA 4202-22 is located on the western edge of the 
survey region. Kohoutek \& Wehmeyer (1999) report H$\alpha$ emission from
the star,  citing ``Dolidze (1971) No. 3'' which probably 
corresponds to Dolidze (1971). This may be where the WR candidacy given in 
SIMBAD comes from although we could not confirm this due to the 
unavailability of 
the Russian
circular, and it is otherwise unnoted in the literature. Figure
\ref{fig.hb_3c} shows a
three-color image from the 3.6, 5.8, and 24 \mum\ data, and the individual
images are in Figure \ref{fig.hb_all}. We cannot
say definitively there is no emission from PAHs, as we do not have the 8
\mum\ data which would contain the (typically) strongest feature. However, 
there is little extended emission in the 5.8 \mum\ data 
that would have contained the 6.2 \mum\ feature if it were present at any 
significant strength. 

\begin{figure}
\plotone{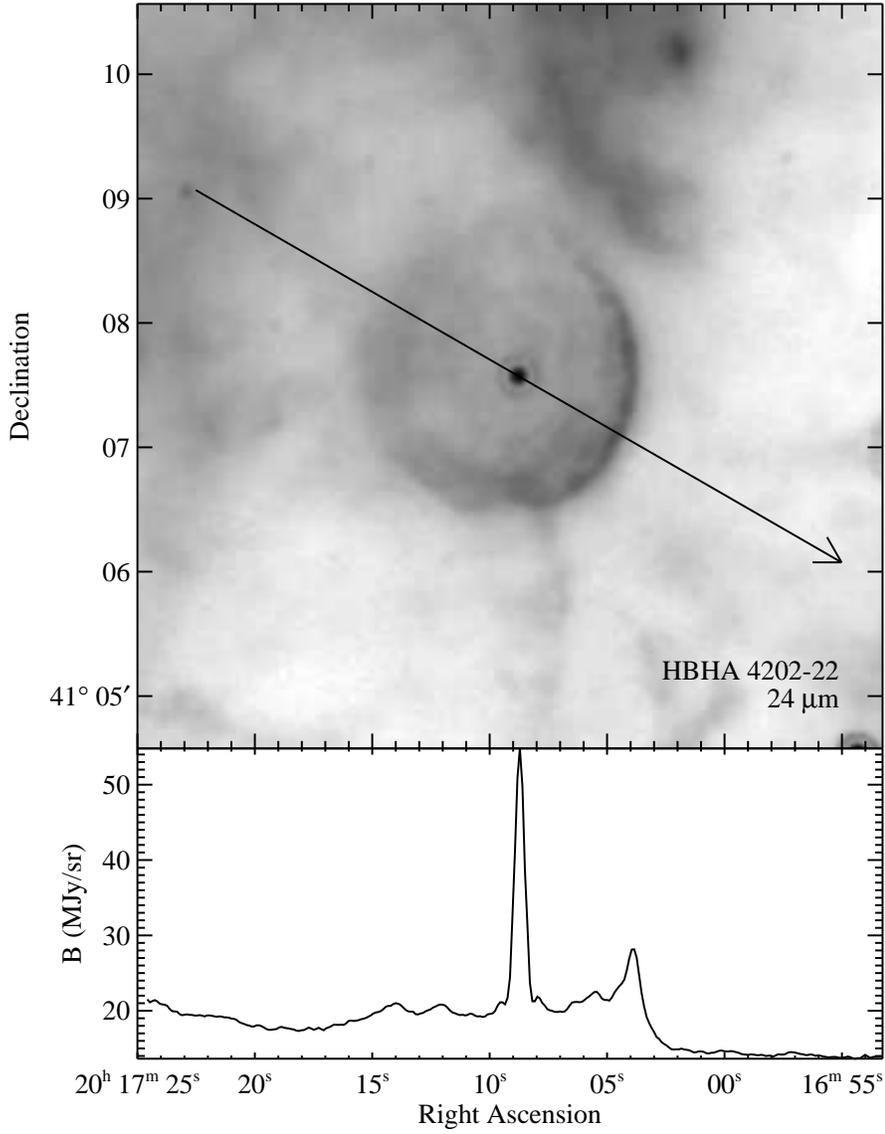}
\caption{(top) The 24 \mum\ ring around HBHA 4202-22. (bottom) Slice through
the 30\arcdeg\ position angle indicated by the line in the top figure. This angle,
which corresponds to the narrowest part of the ring, is just 1\arcdeg\ off of the
31\arcdeg\ proper motion direction. 
}\label{fig.hbslice}
\end{figure}

Figure \ref{fig.hbslice} shows a slice through the ring at position angle 
30$\arcdeg$, which corresponds to the narrowest point in the ring. 
The sharper, southwest edge of the ring is $\sim1\arcmin$ from 
the central star, or about 1.9 pc, assuming a distance of 6.5 kpc (see 
modeling discussion below). The ring is 
more extended to the northeast, $\sim1\farcm2$, or about 2.3 pc,  and the edge 
is less well-defined. Two possible
explanations come to mind. First, the surrounding medium could be less dense
to the northeast, allowing the shell to expand more easily in that direction.
However, the 5.8 \mum\ data indicate that if anything, there is more material 
to the east-northeast than toward the southwest. A second possibility is that
the star is moving toward the southwest and the material on the ``front'' 
side of the shell is being compressed. Indeed, the proper motion of the HBHA 
4202-22 is $\mu_{\alpha}=-4.5$ mas yr$^{-1}$, 
$\mu_{\delta}=-2.7$ mas yr$^{-1}$ (Zacharias et al. 2004), which 
corresponds to a position angle of 31\arcdeg, nicely consistent with the 
compression direction. The distance to this object combined with the proper
motion suggest that it may be a runaway star.

\begin{figure}
\plotone{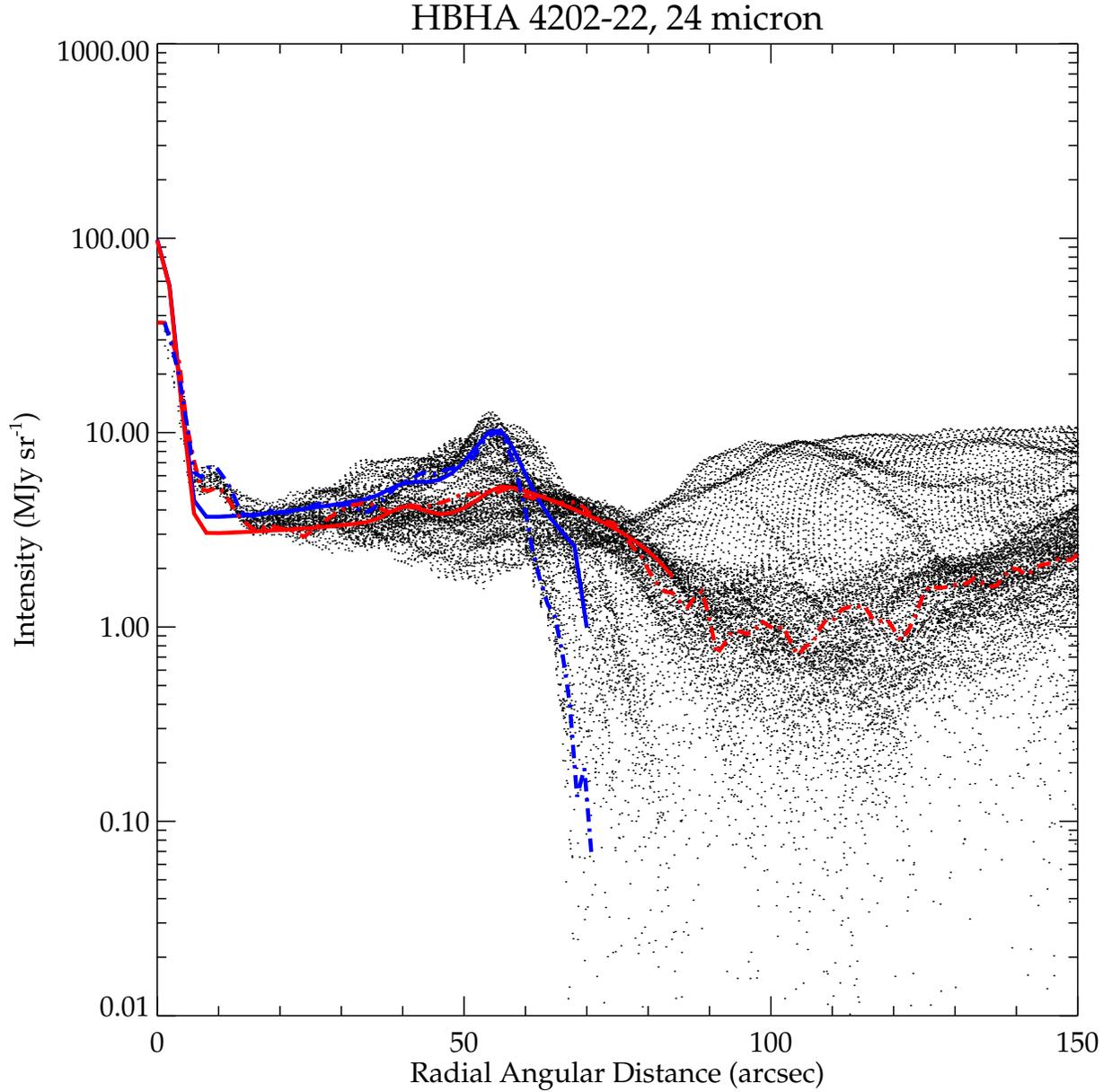}
\caption{Radial profiles for HBHA 4202-22. The dots show the 24 \micron\
data; the dash-dotted lines show a horizontal slice through the data, red for
the extended side, blue for the compressed side. The solid colored lines show
the model results. (A color version of this figure is available in the online
journal.)
}
\label{fig.hbhamodel}
\end{figure}

\begin{figure}
\plotone{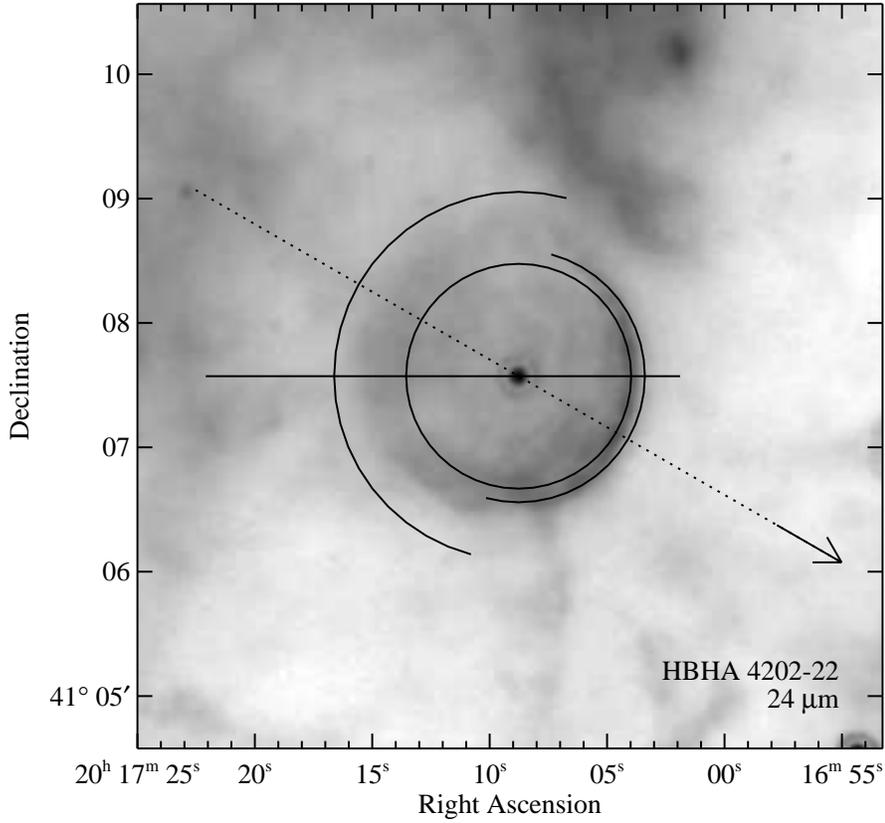}
\caption{Approximate range of the extended and compressed shells. Arcs 
showing the extent of the dust models superposed on the 24 \micron\ image. 
Their angular extent was estimated by eye,
not fit by the model. The solid horizontal line indicates the radii shown with
the red and blue lines in Figure \ref{fig.hbhamodel}. The dotted line with the arrow 
indicates the direction of the 
proper motion of HBHA 4202-22. 
}
\label{fig.hbhaarcs}
\end{figure}

We used a modified version of the radiative transfer code of Egan, Leung, \&
Spagna (1988) to model the dust shells around HBHA 4202-22.   
We modeled the central star as an A0 supergiant with T$_*$=10,000 K, 
L$_*$=$10^5$ \lsun\ at a distance of 6.5 kpc, i.e., well behind the 
Cygnus-X region. While these parameters are formal inputs to the model, 
they, too, were varied in order to get the best fit to the observed data.
As noted above, the shell around HBHA 4202-22 is compressed toward the
southwest compared to the northeast. Therefore, we fit two models to
the two portions. There is also a hint of a ring at $\sim35$\arcsec\
in the 5.8 \micron\ data, so an inner shell is also included in the model.
Table \ref{tab.hbhac} gives the derived parameters using amorphous
carbon, with a particle size of 0.135 \micron, an opacity of
$\tau_{8.035~\micron}=7.5\times10^{-6}$ ($\tau_{0.55~\micron}=
1.77\times10^{-2}$), and grain constants from Mathis \& Whiffen (1988). 
Figures \ref{fig.hbhamodel} and \ref{fig.hbhaarcs} show the results of the 
model compared to the
data for the primary shells (compressed and not) at 24 \micron. Models using 
silicate-rich dust were also considered, which result in cooler dust grains. 
The match to the data was not as good, though, so only the carbon-rich 
results are presented in Table
 \ref{tab.hbhac}.

\subsection{G79.29+0.46}

\begin{figure}
\plotone{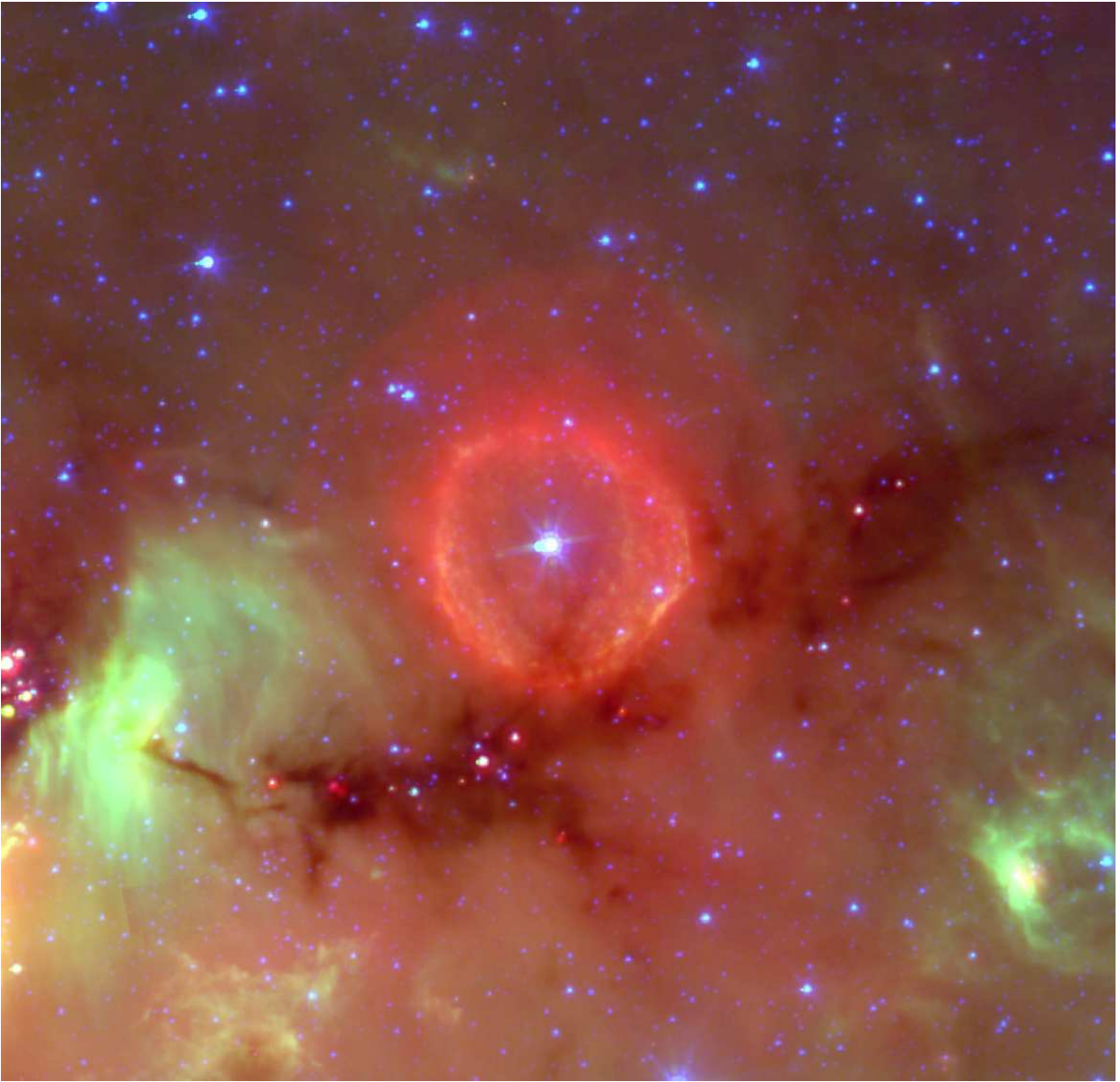}
\caption{Three color image of G79.29+0.46. Red: 24 \mum; green: 8 \mum; blue:
3.6 \mum. The image is $15\arcmin\times15\arcmin$ or $\sim$8.75 pc at a 
distance of 2 kpc, or $\sim$13.13 pc at a 
distance of 3 kpc (see text).
 }
\label{fig.g79_3c}
\end{figure}

\begin{figure}
\plotone{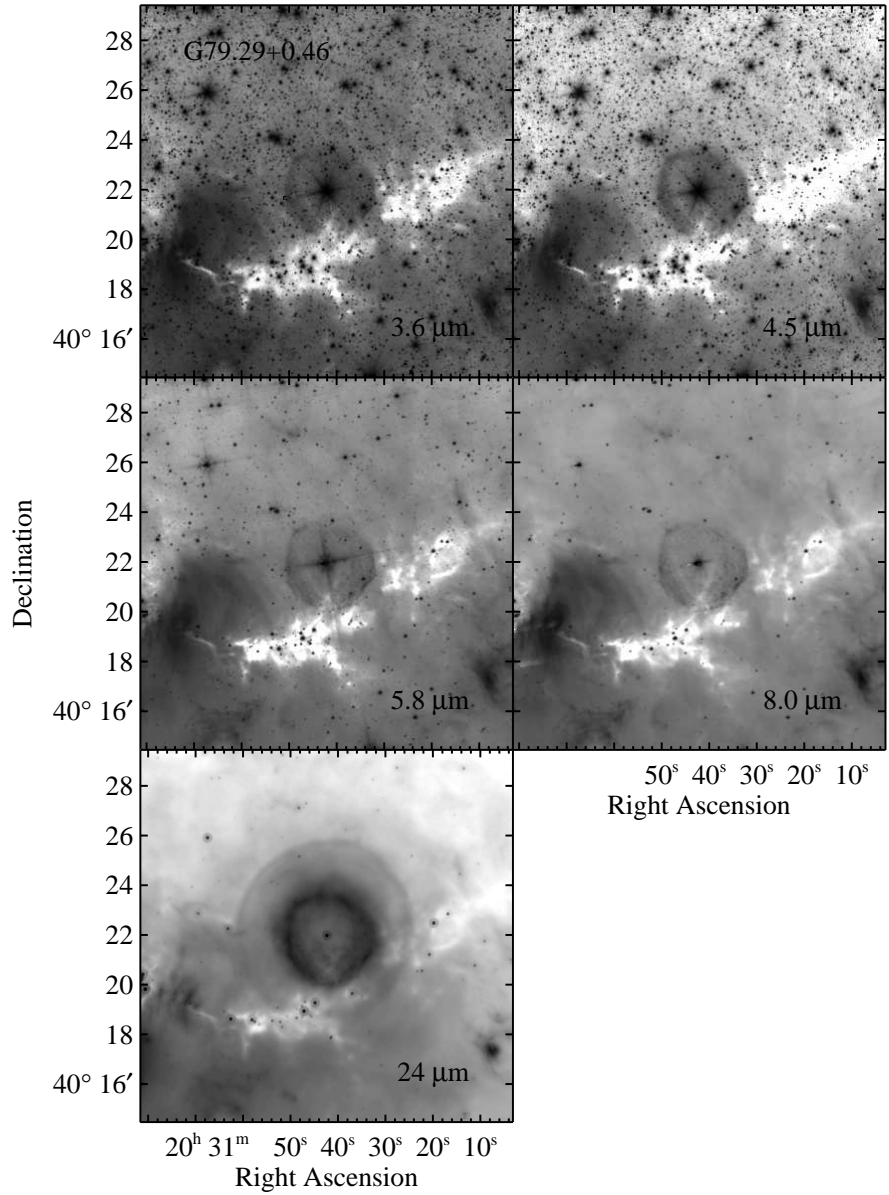}
\caption{Individual images of G79.29+0.46 in the IRAC bands (top two rows) and
at 24 \mum\ (bottom). Note that the gap in the south of the inner ring is
caused by foreground absorption associated with the infrared dark cloud 
stretching across the bottom of the images. Gray scales are logarithmic.
 }
\label{fig.g79_5}
\end{figure}

Figure \ref{fig.g79_3c} shows the three-color image of G79.29+0.46 from the
3.6, 8, and 24 \mum\ data.  It was first found to have an $\sim$4\arcmin\ 
circumstellar ring in the radio by Higgs et al. (1993), who also suggested
that the central object is probably a luminous blue variable (LBV). 
Subsequently, the same ring was detected in the infrared at a number of
wavelengths from 8 to 60 \mum\ with the {\em Infrared Astronomical Satellite}
({\em IRAS}; Waters et al. 1996), the {\em Infrared Space Observatory} ({\em
ISO}; Wendker et al. 1998), and {\em
MSX} (Egan et al. 2002, Clark et al. 2003). This inner ring is detected 
with all four IRAC
bands as well as the 24 \mum\ MIPS band. Figure  \ref{fig.g79_5} shows
the individual images from all five \spi\ bands. The apparent break in the 
south of the ring, particularly in the IRAC bands, is actually due to an 
infrared dark cloud (IRDC) that extends south-southwest in the figures. 
This suggests that as with HBHA 4202-22, G79.29+0.46 may be more distant 
than the 1.7 kpc assumed for the Cygnus-X complex. Interestingly,
though, the CO observations of Rizzo et al. (2008) also show a break in the
CO emission in roughly the same position as the IRDC. Since IRDCs are not known
to absorb molecular line emission in the submillimeter, this suggests that the
two structures might actually be interacting in some fashion. Additionally,
the  infrared ring appears to be somewhat flattened in the southeast, also 
consistent with possible interaction.

Also visible in the 24 \mum\ data in Figures  \ref{fig.g79_3c}  and  
\ref{fig.g79_5} is a newly discovered outer ring $\sim7\arcmin$ across. This 
large ring probably represents an earlier episode of mass loss from G79.29+0.46 
compared to the previously known inner ring.
Additionally, the 24 \mum\ emission from the inner ring is much more extended 
than that from the shorter wavelengths. This kind of structure, where the
IRAC emission ring is interior to a larger round emission structure at 24 \mum,
is occasionally seen in the MIPSGAL catalog of 24 \micron\ rings and disks 
(Mizuno et al. 2010). Usually, the 8 \micron\ emission appears to 
be co-spatial with the 24 \micron\ emission, if present, but most often it is
absent entirely (as may be the case with HBHA 4202-22). 

\begin{figure}
\plotone{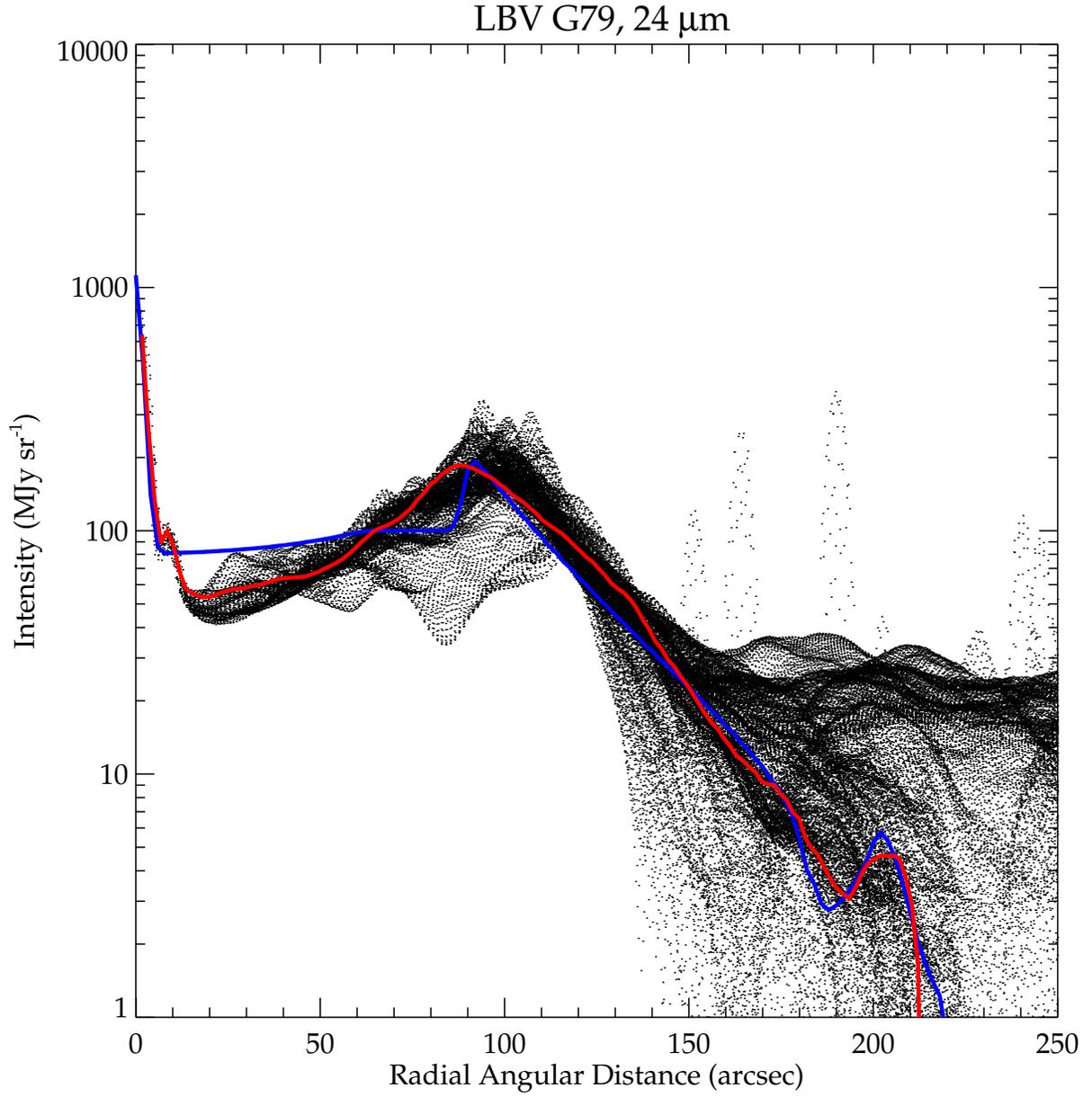}
\caption{Radial profiles of the rings around G79.29+0.46. The red line is a 
slice through the data westward of the central star. The blue line shows the 
model fit. (A color version of this figure is available in the online
journal.)
 }
\label{fig.g79_rad}
\end{figure}

As with HBHA 4202-22, we modeled the dust shells for G79.29+0.46. The stellar
parameters are $T_*=18,000$ K, $L_*=4\times10^5$ \lsun, 
$R_*=4.567\times10^{12}$
cm, with a distance of $d=9.25\times10^{21}$ cm, i.e. slightly behind Cygnus-X 
at $\sim$3 kpc. Table \ref{tab.g79} lists the derived parameters for the inner 
and outer shells. A mix of amorphous carbon grains and PAHs (Li \& Draine
2001) was used in the model, as the dust alone could not produce the flux 
detected in the IRAC bands and the strong PAH emission seen in the
IRS observations of Umana et al. (in preparation)\footnote{Morris et al. 
(2008) present an infrared spectrum  of G79.29+0.46 that included
IRS data as well as ISOSWS and near-infrared spectra. 
Presumably the IRS spectrum is from the central star as it does not show the 
strong PAHs that are detected by Umana et al. toward the shell.}. The 
amorphous carbon grains dominate the mass and opacity, although the (much) 
smaller PAHs are more numerous. While the inner shell is modeled
as a single shell with a density law of $r^{-3.5}$, the 24 \micron\ flux 
profile (Fig. \ref{fig.g79_rad}) suggests that a more complex model with 
a different density law for the 
interior of the shell might give better results.

\section{Summary \label{sec.conc}}

We have detected extended 24 \micron\ emission around three evolved stars
in the Cygnus-X Legacy survey region. A bipolar nebula was found around the
carbon star BD+43 3710. IRS spectra show that the emission is due to small dust
grains, rather than [O IV] line emission or the broad MgS feature. An 
asymmetric ring of 24 \micron\ emission was observed around the Wolf-Rayet 
candidate 
HBHA 4202-22. The ``compressed'' portion of the ring is in the direction of the
projected velocity vector derived from the observed proper motion of the star.
Radiative transfer models imply that the central star is an A-type supergiant
at a distance of $\sim$6.5 kpc and the ring is composed of amorphous carbon 
grains. A large, $\sim7\arcmin$ diameter ring was
detected at 24 \micron\ around the LBV candidate G79.29+0.46, a new addition 
to the 
well-known inner ring. The inner ring was detected at all four IRAC bands as 
well as at
24 \micron. The shells were modeled with a mixture of amorphous 
carbon grains and hot PAHs. An infrared dark cloud blocks part of the ring 
emission at the shorter wavelengths, which along with the model results 
indicates the LBV candidate is at least on the far side of the Cygnus-X 
region, if not further.

While somewhat disparate in their properties, these three objects demonstrate 
the potential of the \spi\ data archives. The thrust of the Cygnus-X Legacy 
project was to explore the richest high mass star formation region within 2 
kpc. Similarly, the larger \spi\ Legacy programs that mapped the inner
Galactic plane with IRAC (GLIMPSE, the Galactic Legacy Infrared Mid-Plane 
Survey Extraordinaire; Benjamin et al. 2003) and MIPS (MIPSGAL; Carey et al. 
2009) also emphasized star formation studies in their proposals and overview 
papers. In each case, though, a wealth of information on circumstellar 
structures around evolved stars has also been obtained.
For example, hundreds of small 24 \mum\ emission 
rings have been found in the MIPSGAL survey data (Mizuno et al. 2010), many of
which are similar to the structures around HBHA 4202-22 and G79.29+0.46 (the 
main difference being that most of the MIPSGAL rings have no detected 24 \mum\
central source). As with our two ring sources, those objects that have previous
identifications are often identified as emission line stars. It is also 
interesting to note that we detect extended emission in the IRAC bands only
toward the inner ring of G79.29+0.46, and the MIPSGAL rings have IRAC
counterparts less than 15\% of the time. Data from follow-up IRS 
observations are currently being analyzed to determine what mechanism is 
responsible for the 24 \micron\ emission in a selection of the MIPSGAL rings 
(Flagey et al. in preparation) - amorphous carbon, as in BD+43 3710; 
[O IV] as in the Morris et al. and Billot et al. supernova remnants; the
as-yet-unidentified 21 \mum\ feature seen in post-AGB stars; or some other 
mechanism. Results from these, as well as other studies underway, will enable 
us to better place our three objects in the broader 
context of evolved stars with circumstellar envelopes in the Galaxy.

\acknowledgements
This work is based in part on observations made with the 
{\em Spitzer Space Telescope}, which is operated by JPL/Caltech
 under NASA contract 1407. Support for this work was 
provided in part by NASA. This research has made use of NASA's 
Astrophysics Data System Bibliographic Services,  data products from 2MASS
 which is a joint project of the University of Massachusetts and IPAC/Caltech
funded by NASA and the NSF,  and the Simbad database 
operated at CDS, Strasbourg, France. We thank the anonymous referee for useful 
suggestions to improve the paper.

{\it Facilities:} \facility{\spi\ (IRAC, IRS, MIPS)}

\clearpage

\begin{deluxetable}{lrrr}
\tablewidth{0pt}
\tablecaption{Evolved Objects in Cygnus-X: Point Source Detections at 24 \mum}
\tablehead{
\colhead{Object} & \colhead{\# in 24 \mum} &
\colhead{\# Det'd at 24 \mum} & \colhead{Detection}\\
\colhead{Type} & \colhead{Region} & \colhead{in Region} & \colhead{Rate} }
\startdata
PNe            &   7 &   5 &  71\%\\
WRs            &   9 &   9 & 100\%\\
Post-AGBs      &   2 &   2 & 100\%\\
Carbon Stars   &  52 &  51 &  98\%\\
S Stars        &   6 &   5 &  83\%\\
SNRs           &  5? &  0? &  0?\%\\
Miras          &   4 &   4 & 100\%\\
semi-regulars  &   4 &   4 & 100\%\\
Cepheids       &   4 &   4 & 100\%\\
Dwarf Novae    &   3 &   1 &  33\%\\
other V*s      &  85 &  61 &  72\%
\enddata
\tablecomments{Object type as reported in Simbad. This is likely to be an
incomplete listing, since not all objects are typed correctly. BD+43 3710,
for instance, is listed merely as a star. There were an additional 40 variable
stars, mostly generic ``V*'' type but including one RR Lyr, which had 
insufficiently precise 
coordinates to determine if they had 24 \mum\ counterparts. It is not
clear how big the SNRs are expected to be at 24 \mum, so the data inspection
may not have been sensitive to their emission. }
\end{deluxetable}

\begin{deluxetable}{lrrrrrr}
\tablewidth{0pt}
\tablecaption{Source Parameters}
\tablehead{
&&&&&\colhead{Spectral} & \colhead{Object}\\
\colhead{Name} & \colhead{RA (J2000)} & \colhead{Dec. (J2000)} & 
\colhead{$l$} & \colhead{$b$}& \colhead{Type} &  \colhead{Type}}
\startdata
\object{BD+43 3710} &  $20^h~45^m~34.^s73$ & +43\arcdeg~32\arcmin~27\farcs28 &
83.3755 &0.3400 & R: & Star \\
\object{HBHA 4202-22} & 20~17~8.12 & +41~07~27.0 & 78.3203 &3.1536 & WR:, A0I & Em. line star \\
\object[NAME G79.29+0.46*]{G79.29+0.46} & 20~31~42.28 & +40~21~59.1 & 80.9005 &1.6435 & B[e] I: & Star
\enddata
\tablecomments{Object types as reported in Simbad. Spectral types are from 
Nassau \& Blanco (1954), Kohoutek \& Wehmeyer (1999) and this work, and Voors et al. (2000) 
for BD+43 3710, HBHA 4202-22, and G79.29+0.46, respectively, although see 
text for discussion.}
\end{deluxetable}

\begin{deluxetable}{lrrrrrrrrrrr}
\rotate
\tablewidth{0pt}
\tabletypesize{\scriptsize}
\tablecaption{Source Fluxes\label{tab.flux}}
\tablehead{
\colhead{Source} & \colhead{B} & \colhead{V} & \colhead{R} & \colhead{J} &
\colhead{H} & \colhead{K} & \colhead{[3.6]} & \colhead{[4.5]} & \colhead{[5.8]}
& \colhead{[8.0]} & \colhead{[24]}
}
\startdata
BD+43 3710 & 11.22$\pm$0.04 mag   &10.06$\pm$0.04 & 9.35$\pm$0.04 & 
6.60$\pm$0.02 & 6.14$\pm$0.02& 5.88$\pm$0.02 & 5.76$\pm$0.02 & 5.63$\pm$0.02&  
5.54$\pm$0.02 &  5.45$\pm$0.02 & 5.09$\pm$0.05\\
     & 0.135 Jy    & 0.334 & 0.537 & 3.66  & 3.57 & 2.96 & 1.40 & 1.01&  
0.70 &   0.42 & 0.07 \\ 
HBHA 4202-22 & 16.54$\pm$0.04 & 15.18$\pm$0.04 & 14.53$\pm$0.04 & 
10.15$\pm$0.02 & 9.27$\pm$0.02 & 8.65$\pm$0.02 & 7.89$\pm$0.02 &\nodata& 
7.20$\pm$0.02 &\nodata& 5.39$\pm$0.05\\
     & 0.001  & 0.003 & 0.005 & 0.139 & 0.202& 0.231& 0.196&\nodata& 
0.151&\nodata &0.050\\
G79.29+0.46  & 19.30$\pm$0.04 & 16.73$\pm$0.04 & 15.20$\pm$0.04 & 
6.91$\pm$0.03 & 5.29$\pm$0.02& 4.33$\pm$0.02\tablenotemark{a}&\nodata&
\nodata& 3.07$\pm$0.02\tablenotemark{a}& 2.86\tablenotemark{a}&
2.05$\pm$0.04\tablenotemark{a} \\
     & 7.9(-5)    &7.2(-4)&2.5(-3)& 2.73  & 7.81 & 12.37&\nodata &
\nodata& 6.82& 4.61& 1.09
\enddata
\tablenotetext{a}{Possibly saturated, so the uncertainties are 
undoubtedly underestimated.}
\end{deluxetable}

\begin{deluxetable}{lll}
\tablewidth{0pt}
\tablecaption{Dust Shell Parameters: HBHA 4202-22\label{tab.hbhac}}
\tablehead{
\colhead{Primary shell} &\colhead{Compressed} & \colhead{Extended}
}
\startdata
Inside radius      & 5.28$\times10^{18}$ cm  & 5.28$\times10^{18}$  cm \\
Thickness          & 0.65$\times10^{18}$ cm & 3.4$\times10^{18}$ cm \\
Dust density law      & $r^{-4}$  & $r^0$ \\
Dust temperature   & 156.0--151.2 K & 156.0--137.4 K \\
Dust opacity       & $\tau_{8.035~\micron}=1.4\times10^{-4}$ \\
                   & $\tau_{0.55~\micron}=1.77\times10^{-2}$ \\\hline
\sidehead{Inner shell}\hline
Inside radius      & 5$\times10^5$ \rstar\\
                   & 3.70$\times10^{18}$  cm \\
Thickness          & 0.70$\times10^{18}$  cm \\
Dust density law      &  $r^{-2}$\\
Dust temperature   & 196.6--188.2 K\\
Dust opacity       & $\tau_{8.035~\micron}=7.5\times10^{-6}$ 
\enddata
\end{deluxetable}

\clearpage

\begin{deluxetable}{lll}
\tablewidth{0pt}
\tablecaption{Dust Shell Parameters: G79.29+0.46\label{tab.g79}}
\tablehead{
\colhead{} &\colhead{Inner Shell} & \colhead{Outer Shell}
}
\startdata
Inside radius      & 3.973$\times10^{18}$ cm  & 8.99$\times10^{18}$  cm \\
                   & 8.7$\times10^5 R_*$      & 1.1 R$_{inner}$ \\
Thickness          & 4.2$\times10^{18}$ cm & 0.5$\times10^{18}$ cm \\
Dust density law      & $r^{-3.5}$  & $r^{-2}$ \\
{\bf Dust temperature}   \\
 am. carbon        & 107.6--89.9 K  & 87.75--86.57 K \\
 PAHs\tablenotemark{a} & 1500--354 K  & 96.1--94.4 K \\
{\bf Dust components}\tablenotemark{b}  \\
$a_{am.car.}$    &  0.45 \micron  & 0.40 \micron \\
$a_{PAHs}$        & 5.0$\times10^{-4}$ \micron  & 5.0$\times10^{-4}$ \micron \\
$N_{am.car.}$& 2\% & 3\% \\
$N_{PAHs}$       & 98\%  & 97\% \\
{\bf Dust opacity}       \\
$\tau_{24~\micron}$ & $1.324\times10^{-4}$ & $9.81\times10^{-6}$\\
$\tau_{8.0~\micron}$ & $2.70\times10^{-3}$ & $2.00\times10^{-4}$\\
$\tau_{0.55~\micron}$ & $1.27\times10^{-2}$ & $9.39\times10^{-4}$
\enddata
\tablenotetext{a}{PAH temperature is non-equilibrium for the inner shell and
declines as $r^{-2}$. The outer shell is an equilibrium temperature.}
\tablenotetext{b}{$a$= grain size; $N$= number density within the shell}
\end{deluxetable}

\clearpage

\end{document}